\documentclass{sigchi}

 \toappear{Technical Report. Service Science And Innovation Departement \\
 {\emph{CRP Henri Tudor}}, September, 2013, Luxembourg, Luxembourg. \\
  \\
}

\usepackage{balance}  
\usepackage{graphics} 
\usepackage{times}    
\usepackage{url}      

\makeatletter
\def\url@leostyle{%
  \@ifundefined{selectfont}{\def\UrlFont{\sf}}{\def\UrlFont{\small\bf\ttfamily}}}
\makeatother
\def\pprw{8.5in}
\def\pprh{11in}

\usepackage[pdftex]{hyperref}
\hypersetup{
pdftitle={SIGCHI Conference Proceedings Format},
pdfauthor={LaTeX},
pdfkeywords={SIGCHI, proceedings, archival format},
bookmarksnumbered,
pdfstartview={FitH},
colorlinks,
citecolor=black,
filecolor=black,
linkcolor=black,
urlcolor=black,
breaklinks=true,
}

\usepackage{color, colortbl}
\definecolor{Gray}{gray}{0.9}

\begin{document}

\title{GENIUS: Generating Usable User Interfaces}

\numberofauthors{2}
\author{
  \alignauthor Jean-S\'{e}bastien Sottet\\
    \affaddr{Public Research Center Henri Tudor}\\
    \affaddr{29 Avenue John F. Kennedy L-1855 Luxembourg}\\
    \email{jean-sebastien.sottet@tudor.lu}\\
  \alignauthor Alain Vagner\\
    \affaddr{Public Research Center Henri Tudor}\\
    \affaddr{29 Avenue John F. Kennedy L-1855 Luxembourg}\\
    \email{alain.vagner@tudor.lu}\\
}

\maketitle

\begin{abstract}
In this report we describe the implementation and approach developed during the GENIUS Project. 
The GENIUS project is about the generation of usable user interfaces. It tries to cope with issues related to automatic generation where, usually end-user complain about the poor quality (in term of usability) of generated UI. 
To solve this issue GENIUS relies on Model-Driven Engineering principles and several MDE tools. Notably, it consists in a set of metamodels specific to the interaction, a set of model transformation embedding usability criteria and an  environment for model execution/interpretation.
\end{abstract}

\keywords{
Engineering Human-Computer Interaction; User Interfaces Generation; Model-Driven Engineering;
}


\section{Introduction}
Early (80's 90's) Model-based approaches mostly focused on the modelling, design and generation of User Interfaces (UI). These approaches were called Model-based User Interfaces (MBUI). MBUI approaches aim at reducing the time of production of UI from, e.g., a functional specification or a data schema. The goal was to  avoid redundant work (between data and UI specification) and to automatise as much as possible the UI implementation process. 

However, such approaches evolved during the last decades. They  promised the reduction of production costs and enhancement of quality by the promotion of User Centred Design~\cite{ISO/IEC13407} processes. MBUI approaches were also unified under a common framework~\cite{Calvary2003}~\footnote{The CAMELEON framework, which is also addressing the adaptability of UI, defines the context of use which is not explicitly quoted in this article.} and gave rise to many standards of modelling languages at the W3C such as~\cite{UsiXMLW3C}.

MBUI approaches, even if relying on advanced and long term research, are missing some qualities, notably when considering automatic and semi-automatic generation. That is, the generated UI are perceived of poor quality and of bad usability. Moreover, recent works \cite{Sottet2007, Fernandez2009, Frey2011a}  show the importance of considering the usability in the transformation/generation step of UI Models.

\section{The GENIUS Modelling Framework}
The objective behind the GENIUS methodology is to improve the quality (principally the usability) of automatically generated UI. The framework we depict here aims at better UI quality and thus orienting the concerns of the generation/transformation to the quality of produced results.  We evaluated the relevance of our approach on a real case study for reporting and documenting construction projects. The case study used here, implements a web version of an existing application, CRTI-WeB~\footnote{http://info.crti-web.lu}, used by professionals of the construction sector.

\subsection{Initial Generation Framework}
\begin{figure}[htbp]
\centering
\includegraphics[width=0.2\columnwidth]{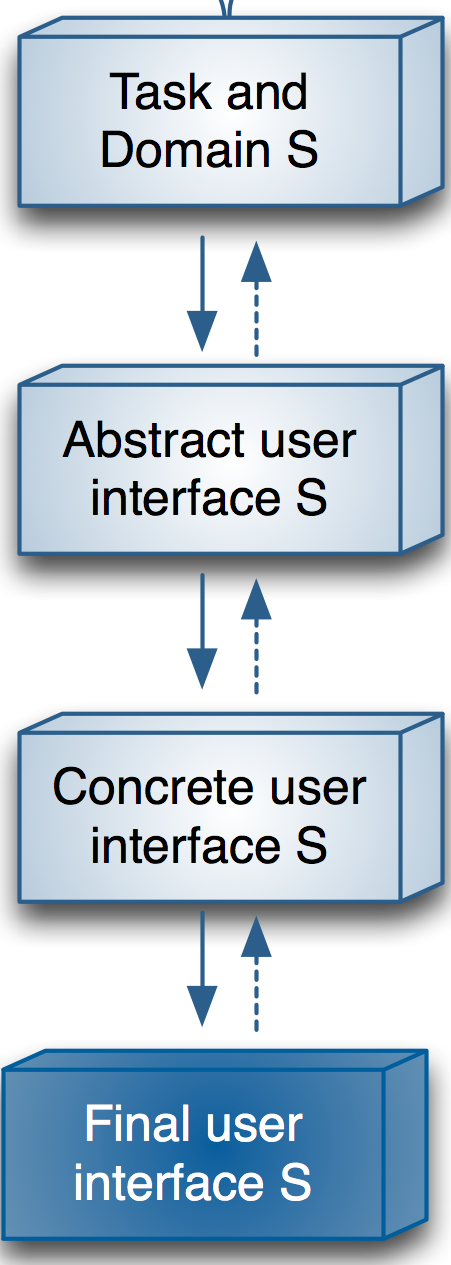}
\caption{Models and transformations as presented in the CAMELEON reference framework (extracted from CAMELEON specifications).}
\label{fig:cameleon}
\end{figure}

In terms of metamodels usage, GENIUS is  partly compliant with the standard CAMELEON reference framework  \cite{Calvary2003}, see Figure~\ref{fig:cameleon}. CAMELEON is composed of multiple abstraction of UI and interaction. The highest level of abstraction consists in the modelling of the tasks~\cite{Paterno1997} (interaction model) and the domain (i.e. the manipulated domain concepts and their attributes). Then, this model is reified into a lower level of abstraction called abstract user interface (AUI). AUIs are interface models abstracted from any interactor (i.e. buttons, text field, combo-box,etc.) and can be considered as the general interaction flow. The next level of abstraction is the concrete user interface (CUI) which instantiates the interactors. It represents a complete interface but is independent from the computing platform (e.g., smartphone, tablets, pc an the different operating systems). The actual execution of the interface is represented by the lowest level of abstraction, the final user interface (FUI). 

Our framework is roughly the same: task and domain models are still to be modelled whereas only the navigation (workflow) part of AUI is kept into a state chart model. The CUI remains the same. 
However, in order to do automatic generation we need all relevant information to be present in the initial models, that are Task and Domain models. The interaction types (also refers to as canonical abstract types~\cite{Constantine2003a}) are crucial for the generation: they depict the kind of action that user as to do in order to reach its goal using the application. Interaction types are associated with AUI model elements and notably help in finding the right interactor (belonging to CUI model). Thus we associate to each task the interaction type (also called AUI type). 

Transformation from Task model to AUI model can be complex to write: it may be composed of a lot of transformations rules, such as in~\cite{Limbourg2009}.
However, it consists mainly in a rewriting of a concurrent task tree (tree structure + operator) into a nested-graph structure (i.e., container and containment) where container can be identified as~\cite{Luyten2003}. 

Our claim is that after the creation by potentially multiple stakeholders and several refinements, the ``input'' model, is sufficiently complete to generate an executable UI (or as we will see later a set of executable models). In order to have all the required information for generation, we need to build a composed model containing Task, Domain and some AUI information called TDA.

\begin{figure}[htbp]
\centering
\includegraphics[width=\columnwidth]{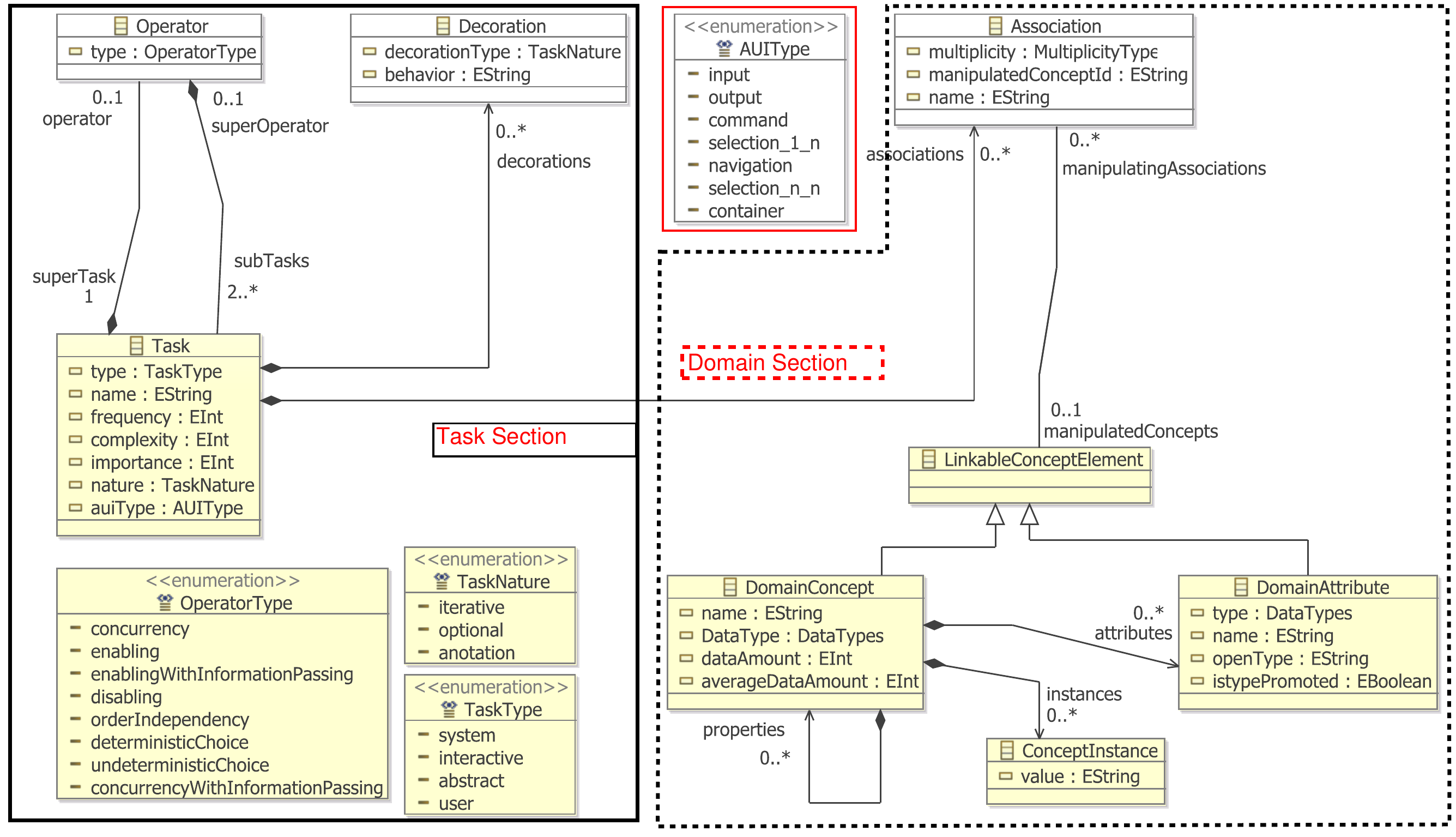}
\caption{TDA - Task Domain Abstraction. Initial Metamodel}
\label{fig:TDA}
\end{figure}

TDA is derived from the notion of ``concrete task model''~\cite{Baron2002} which promotes the enhancement of standard task model with additional information. The task metamodel is an adaptation of the MAD* notation~\cite{Gamboa-Rodriguez1997}, the domain metamodel is adapted from UML class/object diagrams. The Abstract User Interface is not represented as a set of metamodel elements but as additional information standing on each tasks.  Such information gives insight on the interactive nature of the task. This nature can be input/output, selection among n elements, command, container, etc.  All these elements are sufficient to generate a first set of models.
The initial metamodel (see Figure~\ref{fig:TDA}), called TDA (Task Domain Abstraction) contains:
\begin{itemize}
\item from Task model: structural information, that are container/containment relations?  temporal information: what is preceding what?
\item from Domain model: what are the domain concept manipulated during the interaction processes and sub-processes?
\item from AUI model: what is the intended interaction of the task : what does the user actually need to complete his tasks (e.g., selecting one element in a list, etc.)
\end{itemize}

\begin{figure*}[htbp]
\centering
\includegraphics[width=1.8\columnwidth]{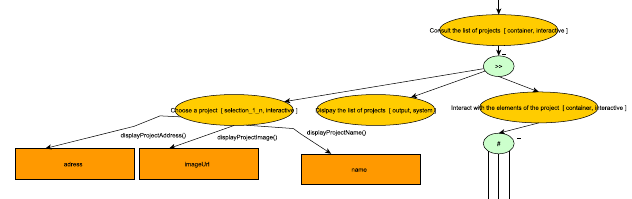}
\caption{Task model: selection of a project Tasks are the ovals. Square are representing concept manipulated by tasks. Labels between the brackets are the AUI Type. Model displayed with yEd Graph Editor}
\label{fig:task}
\end{figure*}

The Figure~\ref{fig:task} is representing a portion of a task for selecting a project and interacting with the selected construction project.

\subsection{Executable Models}
The first set of models that can be obtained from TDA are the Concrete User Interface Model and a State-Chart Model (representing the navigation of the UI).  These models are intended to be executable (see further section on the executable environment)

The Concrete UI (CUI) is quite similar to the ones proposed in the literature and currently being standardised by the W3C\footnote{http://www.w3.org/2005/Incubator/model-based-ui/wiki/Concrete\_User\_Interface\_Model}. The State-Chart is a technical model which depicts all the navigation paths between UI elements (the interaction work-flow). Notably it overcomes the lacks caused by the tree structure of Tasks models by modelling the return paths, multiple direction paths, etc. State-Chart Model is coupled with the CUI model by reference, i.e. a state is aware of which CUI element(s) it relies on. For example clicking on this button leads, by following the state chart transformation, to that window. This model is used here for representing the dynamic aspects of UI which has been identified as an need to enhance standard MBUI approaches~\cite{Vanderdonckt2008}.

\begin{figure}[htbp]
\centering
\includegraphics[width=0.6\columnwidth]{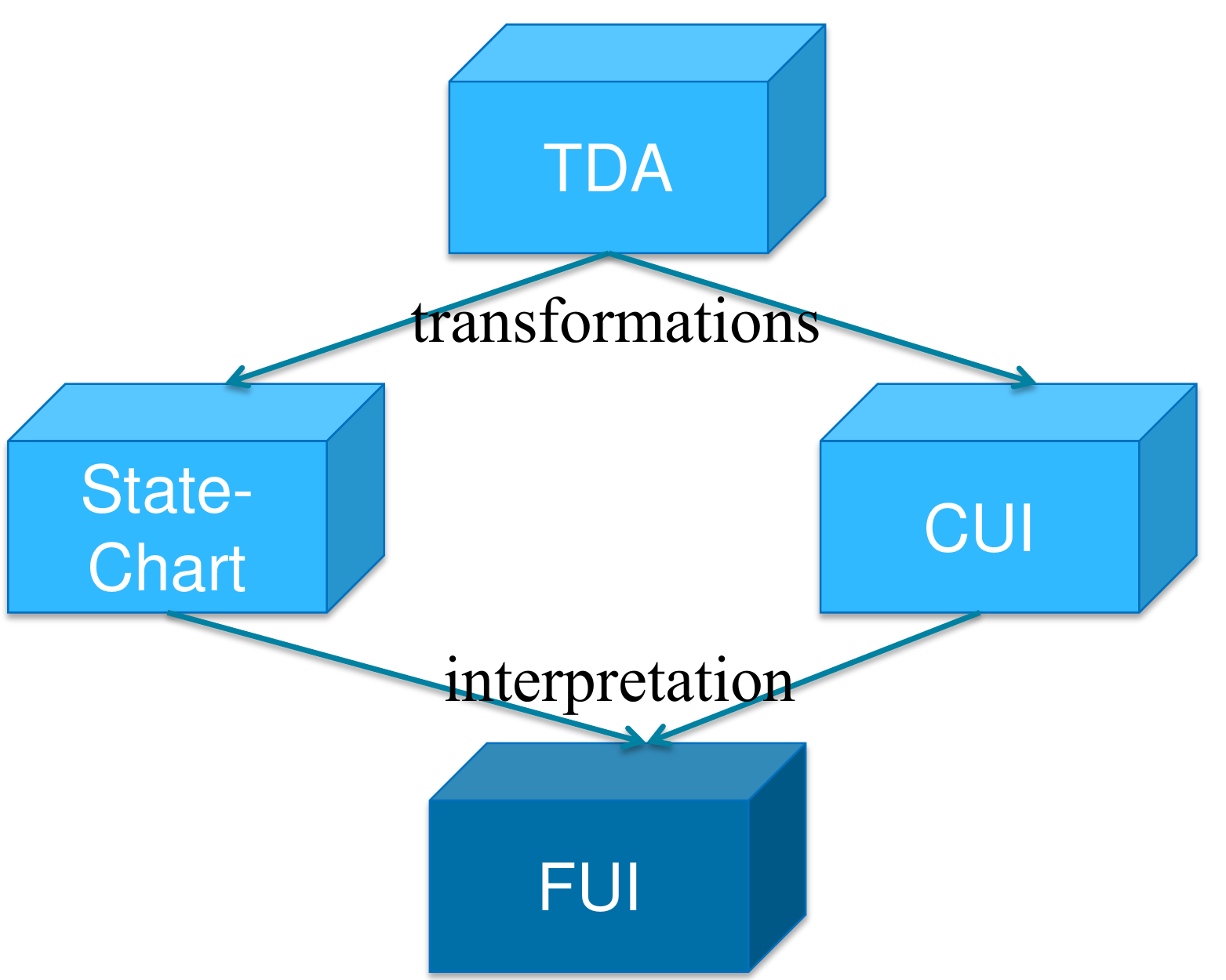}
\caption{The Genius Modelling Framework and transformation Chain.}
\label{fig:genius}
\end{figure}

In Figure~\ref{fig:genius} we depict our modelling framework including transformations (see next sections) for the input model (TDA)  to State-Chart and CUI models. It also includes an interpretation phase (during execution) of CUI and State-charts models.

Regarding the traditional CAMELEON process and the many implementations of it, we go here for simplicity. We deal with the essence of the models: the analysis and design part for the TDA model and the ``executability'' for CUI and State-Chart models. Nevertheless this (apparent) simplicity implies that we put the complexity elsewhere: notably on transformations.


\section{Model-Transformation \& Ergonomic Improvement}
\label{sec:transformation}
The main originality of the GENIUS approach stands on the model transformation. Transformations are usually considered as black-boxes or hidden processes, however they are the key of the UI generation. Our claim is to consider transformations as first order citizen.
We are thus to redefine the consideration of transformations and the principle underlying the transformations. Notably, if we want to address the challenge of generating (more) usable user interfaces, we need to make transformations aware of usability needs.
All the transformation rules are designed with usability designed \textit{a priori} and evaluated \textit{a posteriori}. The \textit{a posteriori} evaluation is made by analysing (alike classical usability evaluation made by usability expert) the result of the transformation chain, on the generated FUI. 

\begin{table}[htbp]
	\begin{tabular}{|p{1.2cm}|p{1.5cm}|p{2cm}|p{2cm}|}
		\hline 
		\rowcolor{Gray}
		 Name & Description & Positive Contribution & Negative Contribution \\ 
		\hline Add filter on List & Add filter within a list with more than 5 elements & 6. Consistency & 2. Workload: 2.1.Brevity  \\ 
		\hline  Hiding Password fields & hide password during the entry & 6.Consistency
		8.Compatibility
		+ Safety
		 & 5.Error management
		5.1.Error protection
		5.3.Error correction \\ 
		\hline 
	\end{tabular} 
\caption{Example of informal transformation description with ergonomic description}
\label{table:informal}
\end{table}

\subsection{Designing usable transformation: usability a priori}
Transformation are, in our approach, to be placed as the same level as modelling practices: transformation are models~\cite{Bezivin2006}. In spite of the equal importance of model and transformations, transformations should not be changed as much as models. Designing a new application implies to build new model(s) but maybe not to modify/create any transformation.  We keep in mind that transformations must keep some generic aspects in order to get benefits from using such model-driven approaches e.g., reusing them many times in different contexts.

Based on the principles of \cite{Sottet2007a} and the statement above, we assert that:
\begin{itemize}
\item \textbf{Assertion 1:} transformations must be exposed and controlled by the designers, up to a certain level regarding their competences and knowledge.
\item \textbf{Assertion 2:} transformations must be aligned with usability properties and thus should embed usability rules.
\item \textbf{Assertion 3:} transformation design must be considered as an iterative process. Transformation are a mean to capitalise knowledge. Transformation design is part of the design process itself. 
\end{itemize}

For the executable part of the transformation we retain for technical effectiveness an existing MDE transformation language (for example ATL).  However as stated in \cite{Panach2011} defining transformations with such a language is a matter of model-driven specialists and not trivial for other kind of stakeholders. Thus, we need to provide new kinds of representations for transformations, called trans-HCI~\cite{Sottet2008} \footnote{The Authors have currently a paper under submission on ``giving control on model-transformation to user experience specialist'' thanks to a specific HCI}.
This kind of specific UI/representation of a transformation is dedicated to specific jobs and roles in the design process.

\begin{figure*}[htbp]
\centering
\includegraphics[width=2\columnwidth]{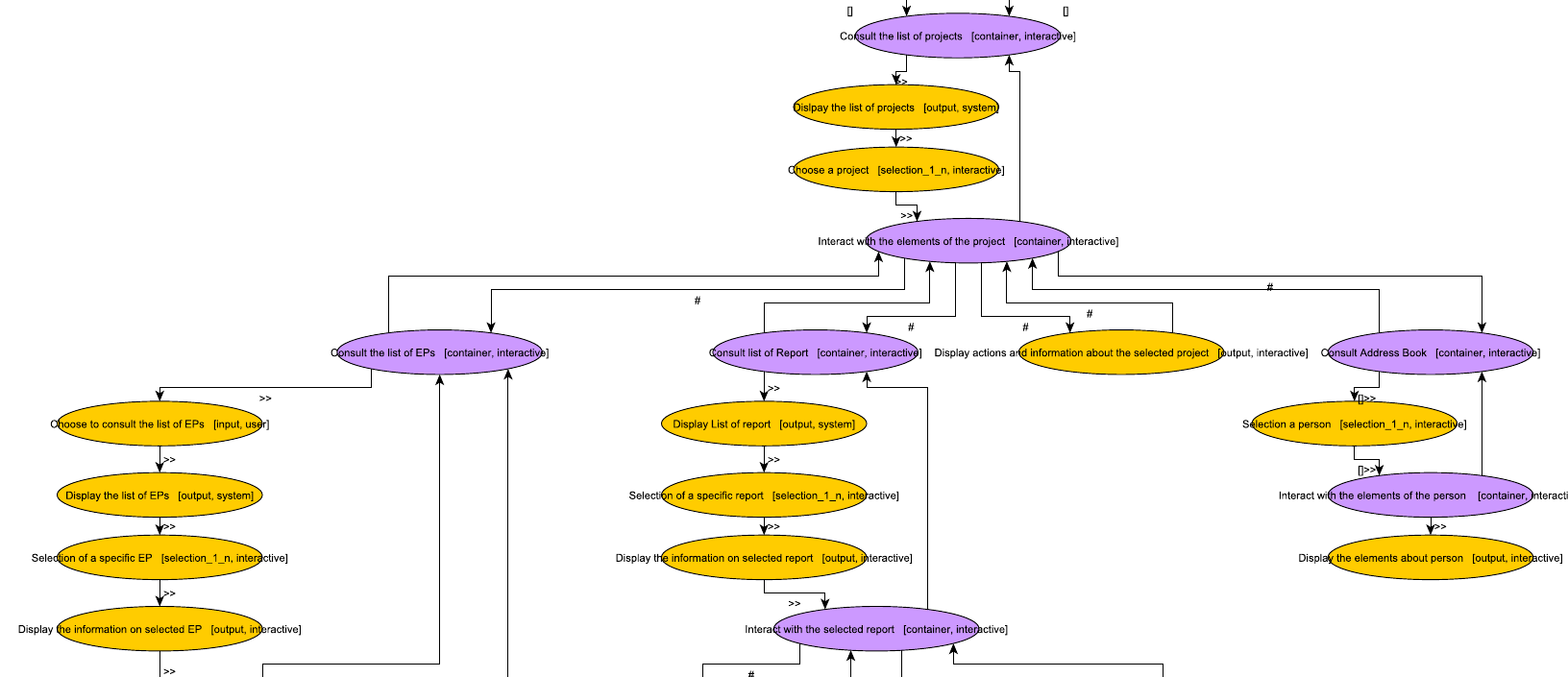}
\caption{Start-Chart model: selection of project and interaction with selected project elements . Model displayed with yEd Graph Editor}
\label{fig:statechartglobal}
\end{figure*}

In executable transformation (e.g., transformations designed with a transformation language) are defined along with usability properties they should fulfil. In this current work we have chosen a particular ergonomic framework defined in~\cite{Scapin1997}. This framework is an analytical tool that help in criticizing existing projects. However, it can be applied to guide software engineering in design situations as shown in \cite{Sottet2007}.

The objective is to finalise a set of transformation rules used together for UI generation. Each rule will be identified and annotated with the ergonomic properties it conveys.

\subsection{Improvement: usability a posteriori}
After this first step (i.e.,defining transformations), we also consider usability defined \textit{a posteriori}: usability  perceived on transformed UI. Thanks to user tests, usability expert evaluates the transformation production. Then the needed usability (e.g., users need guidance because they are lost in their interaction path) is depicted as a requirement. The designers cope with this issue by defining a new transformation (e.g., a transformation function that is inserting breadcrumb on each web-pages). The overall process of usability/ergonomy improvement is more widely depicted in~\cite{Montecalvo2011}.

\section{Model Execution Environment}
\label{sec:execution}

During the GeNIUS project we have developed a web run-time environment capable of executing (i.e., create executable HTML/jQuery user interfaces) state charts and CUI models. Then it executes this web user interface and traces all the actions done by the user(s). This runtime environment is in fact a single page web application  and as such is mainly constituted of web technologies (HTML, CSS, JavaScript). As we mentioned, it is based on the jQuery Mobile framework , which targets mobile platforms. It only requires a small server-side component for the logging of user actions.

We have chosen jQuery mobile because it is a declarative way for HCI design: the designer creates an html structure based on an extension of the HTML5 language (using existing HTML5 extension points) and then the jQuery Mobile library decorates the DOM accordingly. It is noteworthy that this kind of transformation is dynamic. If the application changes some DOM object attributes, jQuery Mobile propagates the changes on the fly, thus enabling us to have an approach based on models at runtime.

The initialization process (creation of the HTML/jQuery pages) use as inputs  CUI and State-Chart Models. On these models we apply template definitions (i.e., can be named transformation rule from CUI to FUI - html -) and some manual extensions (e.g., adding specific color for specific component). This mix of approaches: automatic transformation of CUI models into an HTML page allows for a better management of UI quality. Indeed, if a manual modification become a generic rule, we transfer it to the other transformation rules. Higher is the transformation, for instance from TDA to CUI, more generic and deeper will be the manual modification.

The role of the state machine is to deal with states, tasks and web pages changes. This has to be done in conjunction with the ``runtime context'' (i.e., data stack) which contains all state information to be kept during the transitions. For example if the user select one element (in a selection state), this select element must be shared by the next state. Moreover the state machine handles different types of states depending on the types of associated tasks: window, system and user. Window are of special type since they directly reference a portion of the CUI model encompassing a window content.
Regarding the manual extensions, they are loaded for each state in the state machine and is able to extend the generated user interfaces.

The execution of the application starts when we start the state machine. This machine processes each window state with the following algorithm:

While(current state is not a final state)
\begin{enumerate}
\item For the current (window) state, select the associated CUI elements (contained by a window).
\item Transformation of the CUI elements (contained by  window) compositing into a FUI window. This transformation is made of two steps:
	\begin{enumerate}
	\item Each CUI element to be displayed (i.e., elements of a window) is transformed by the template into FUI.
	\item the template system: JSRender~\footnote{https://github.com/BorisMoore/jsrender} is based on jQuery Templates. The template system produces HTML code compliant with jQuery Mobile that is directly injected in the DOM of the current page.
	\end{enumerate}
\item Loading of the associated data into the templates (Mapping by domain concepts (declared in Json) manipulated by CUI into the FUI)
\item Execution of the manual extensions for the current (window )state.
\item Display of the finalized window and its content to the user
\item Handling of the users action and the information about the corresponding CUI and SC concepts.
\item If FUI element linked to the transition is actuated, search of the next (window) state through the transitions. The state machine generates a (new) window only for window states. So it is necessary to know which types of states are reachable in the frame of a window. 
\begin{itemize}
\item Evaluation of the state preconditions
\item Search of the next window state
\end{itemize}
\end{enumerate}
EndWhile()

\section{Conclusion}
Based on previous researches addressing model-based user interface design and existing modelling frameworks we have defined a refined framework. Alike existing model-based frameworks, it is based on four design models such as in~\cite{Calvary2003} (i.e., task, domain, AUI, CUI).
However we break the standard approach replacing the AUI model by a UI dynamics model (state-chart) that also cope with standard task tree issues. Moreover, this approach allows for dynamic interpretation of model: CUI models are depicting what is actually displayed (i.e., CUI interpreted into FUI) and what are the possible interaction paths (``execution'' of the State-chart model).

This framework puts forward some principle underlying automatic generation putting the model-transformation as first order citizen: transformations are no longer to be seen as  black-boxes. Such transformations are thus not only considering technical aspects but also must cover some quality properties such as usability. In our implementation we propose a usability driven definition of model transformations involving end-users (in tests). Such approach provided us with a repository of usability-aware transformations which helps in generating better quality UI.

\section{Acknowledgements}
This work had been partially supported by the FNR CORE Project GENIUS (C09/IS/13).
We would like to frankly thank Eric Montecalvo for the first implementation of the web runtime environment and the
fruitful conversation about MDE integration.
%
%
%
%
%
\balance

\bibliographystyle{acm-sigchi}
\bibliography{MoDEL,GeniusBiblio}

\begin{thebibliography}{10}

\bibitem{Baron2002}
Baron, M., and Girard, P.
\newblock Suidt: A task model based gui-builder.
\newblock {\em Proceedings of the First International Workshop on Task Models
  and Diagrams for User Interface Design\/} (2002), p 64--71.

\bibitem{Bezivin2006}
{B\'ezivin}, J., {B\"uttner}, F., {Gogolla}, M., {Jouault}, F., {Kurtev}, I.,
  and {Lindow}, A.
\newblock Model transformations? transformation models!
\newblock In {\em Proceedings of the 9th International Conference, MoDELS
  2006}, vol.~4199 of {\em Lecture Notes in Computer Science}, Springer Verlag
  (Berlin, 2006), 440--453.

\bibitem{Calvary2003}
Calvary, G., Coutaz, J., Thevenin, D., Limbourg, Q., Bouillon, L., and
  Vanderdonckt, J.
\newblock A unifying reference framework for multi-target user interfaces.
\newblock {\em Interacting with Computers 15}, 3 (2003), 289--308.

\bibitem{Constantine2003a}
Constantine, L.~L.
\newblock Canonical abstract prototypes for abstract visual and interaction
  design.
\newblock In {\em Interactive Systems. Design, Specification, and
  Verification}. Springer, 2003, 1--15.

\bibitem{Fernandez2009}
Fernandez, A., Insfran, E., and Abrah{\~a}o, S.
\newblock Integrating a usability model into model-driven web development
  processes.
\newblock In {\em Web Information Systems Engineering-WISE 2009}. Springer,
  2009, 497--510.

\bibitem{Frey2011a}
Frey, A.~G., C\'eret, E., Dupuy-Chessa, S., and Calvary, G.
\newblock {QUIMERA}: a quality metamodel to improve design rationale.
\newblock In {\em Proceedings of the third {ACM} {SIGCHI} Symposium on
  Engineering Interactive Computing Systems ({EICS} 2011)}, {ACM} Press (2011),
  265--270.

\bibitem{Gamboa-Rodriguez1997}
Gamboa-Rodriguez, F., and Scapin, D.~L.
\newblock Editing {MAD}* task descriptions for specifying user interfaces, at
  both semantic and presentation levels,.
\newblock In {\em in Proceedings {DSV-IS} '97, 4th International Eurographics
  Workshop on Design, Specification, and Verification of Interactive Systems,}
  (June 1997).

\bibitem{ISO/IEC13407}
ISO/IEC13407.
\newblock Human-centred design processes for interactive systems,, 1999.

\bibitem{Limbourg2009}
Limbourg, Q., and Vanderdonckt, J.
\newblock Multipath transformational development of user interfaces with graph
  transformations.
\newblock In {\em Human-Centered Software Engineering}, A.~Seffah,
  J.~Vanderdonckt, and M.~Desmarais, Eds., Human-Computer Interaction Series.
  Springer London, 2009, 107--138.

\bibitem{Luyten2003}
Luyten, K., Clerckx, T., Coninx, K., and Vanderdonckt, J.
\newblock Derivation of a dialog model from a task model by activity chain
  extraction.
\newblock In {\em Interactive Systems. Design, Specification, and
  Verification}, J.~Jorge, N.~Jardim~Nunes, and J.~Falcão~e Cunha, Eds.,
  vol.~2844 of {\em Lecture Notes in Computer Science}. Springer Berlin
  Heidelberg, 2003, 203--217.

\bibitem{Montecalvo2011}
Montecalvo, E., Vagner, A., and Gronier, G.
\newblock Proposal of a usability-driven design process for model-based user
  interfaces.
\newblock In {\em Proc. of the 2nd Inter. Workshop on USIXML} (2011).

\bibitem{Panach2011}
Panach, J.~I., Pastor, O., and Aquino, N.
\newblock A model for dealing with usability in a holistic model driven
  development method.
\newblock In {\em Proc of 2nd Inter. workshop on USIXML} (2011).

\bibitem{Paterno1997}
Patern{\`o}, F., Mancini, C., and Meniconi, S.
\newblock Concurtasktrees: A diagrammatic notation for specifying task models.
\newblock In {\em Proceedings of the IFIP TC13 Interantional Conference on
  Human-Computer Interaction}, vol.~96 (1997), 362--369.

\bibitem{Scapin1997}
Scapin, D., and Bastien, J.
\newblock {\em Ergonomic criteria for evaluating the ergonomic quality of
  interactive systems}, vol.~16.
\newblock Behaviour \& Information Technology,, 1997.

\bibitem{Sottet2008}
Sottet, J.-S.
\newblock {\em M\'ega-IHM : Mall\'eabilit\'e des Interfaces Homme-Machine
  dirig\'ees par les mod\`eles}.
\newblock PhD thesis, Laboratoire d'Informatique de Grenoble (LIG), Université
  Joseph Fourier., 2008.

\bibitem{Sottet2007}
Sottet, J.-S., Calvary, G., Coutaz, J., and Favre, J.-M.
\newblock A model-driven engineering approach for the usability of user
  interfaces.
\newblock In {\em Proceedings of Engineering Interactive Systems 2007} (2007),
  140--157.
\newblock Joint conferences of {IFIP} {WG2}.7/13.4 10th Conference EHCI, {IFIP}
  {WG} 13.2 1st Conference HCSE , and the 14th Conference {DSV-IS} , University
  of Salamanca, Spain, March 22-24, 2007,.

\bibitem{Sottet2007a}
Sottet, J.-S., Ganneau, V., Calvary, G., Coutaz, J., Favre, J.-M., and
  Demumieux, R.
\newblock Model-driven adaptation for plastic user interfaces.
\newblock In {\em Proc. {INTERACT} 2007, the eleventh {IFIP} {TC13}
  International Conference on Human-Computer Interaction} (2007), 397--410.
\newblock Springer {LNCS} (Lecture Notes in Computer Science), Brasil,
  September 10-14, 2007.

\bibitem{Vanderdonckt2008}
Vanderdonckt, J.
\newblock Model-driven engineering of user interfaces: Promises, successes,
  failures, and challenges.
\newblock {\em In Proc. of 5th Annual Romanian Conf. on Human–Computer
  Interaction ROCHI'2008\/} (2008), 1--10.

\bibitem{UsiXMLW3C}
{W3C}.
\newblock Usixml user interface extensible markup language.
\newblock Tech. rep., W3C MBUI Incubator Group, 2009.

\end{thebibliography}
\end{document}